\documentclass[letterpaper, 10 pt, conference]{ieeeconf}  

\IEEEoverridecommandlockouts                              

\usepackage{graphics} 
\usepackage{epsfig} 
\usepackage{mathptmx} 
\usepackage{times} 
\usepackage{amsmath} 
\usepackage{amssymb}  
\usepackage{booktabs} 
\usepackage{graphicx}
\usepackage{multicol}
\usepackage{svg}
\PassOptionsToPackage{prologue,dvipsnames}{xcolor}
\usepackage{url}

\title{\LARGE \bf
Snap, Segment, Deploy: A Visual Data and Detection Pipeline for Wearable Industrial Assistants
}

\author{Di Wen$^{1}$, Junwei Zheng$^{1}$, Ruiping Liu$^{1}$, Yi Xu$^{1}$, Kunyu Peng$^{1,\dagger}$, Rainer Stiefelhagen$^{1}$
\thanks{$^{\dagger}$ Corresponding author}
\thanks{All authors are with the Computer Vision for Human-Computer Interaction Lab (cv:hci), 
Institute for Anthropomatics and Robotics (IAR), 
Karlsruhe Institute of Technology, 76131 Karlsruhe, Germany. 
Emails: \texttt{\{di.wen, junwei.zheng, ruiping.liu, kunyu.peng, rainer.stiefelhagen\}@kit.edu},\texttt{yi.xu@student.kit.edu}}%
}

\begin{document}

\maketitle
\thispagestyle{empty}
\pagestyle{empty}

\begin{abstract}
Industrial assembly tasks increasingly demand rapid adaptation to complex procedures and varied components, yet are often conducted in environments with limited computing, connectivity, and strict privacy requirements. These constraints make conventional cloud-based or fully autonomous solutions impractical for factory deployment. This paper introduces a mobile-device-based assistant system for industrial training and operational support, enabling real-time, semi-hands-free interaction through on-device perception and voice interfaces.
The system integrates lightweight object detection, speech recognition, and Retrieval-Augmented Generation (RAG) into a modular on-device pipeline that operates entirely on-device, enabling intuitive support for part handling and procedure understanding without relying on manual supervision or cloud services. 
To enable scalable training, we adopt an automated data construction pipeline and introduce a two-stage refinement strategy to improve visual robustness under domain shift. Experiments on our generated dataset, \textit{i.e.}, Gear8, demonstrate improved robustness to domain shift and common visual corruptions.
A structured user study further confirms its practical viability, with positive user feedback on the clarity of the guidance and the quality of the interaction. These results indicate that our framework offers a deployable solution for real-time, privacy-preserving smart assistance in industrial environments. We will release the Gear8 dataset and source code upon acceptance.
\end{abstract}

\section{INTRODUCTION}
\label{sec:intro}

The complexity of manufacturing assembly tasks is escalating, driven by the increasing diversity of components, intricate assembly procedures, and the growing need for customization in contemporary production settings~\cite{c1}.
While full automation is still unfeasible in many contexts, newly onboarded or rotating workers often struggle with steep learning curves, resulting in extended training periods and a higher incidence of assembly mistakes~\cite{c2}.
To address these challenges, vision-based assistance systems have emerged as a promising solution, offering real-time, step-by-step guidance that helps reduce both task completion time and error rates~\cite{c3}.

\begin{figure}[thpb]
\centering
\framebox{\parbox{3in}{
\includegraphics[width=1.0\linewidth]{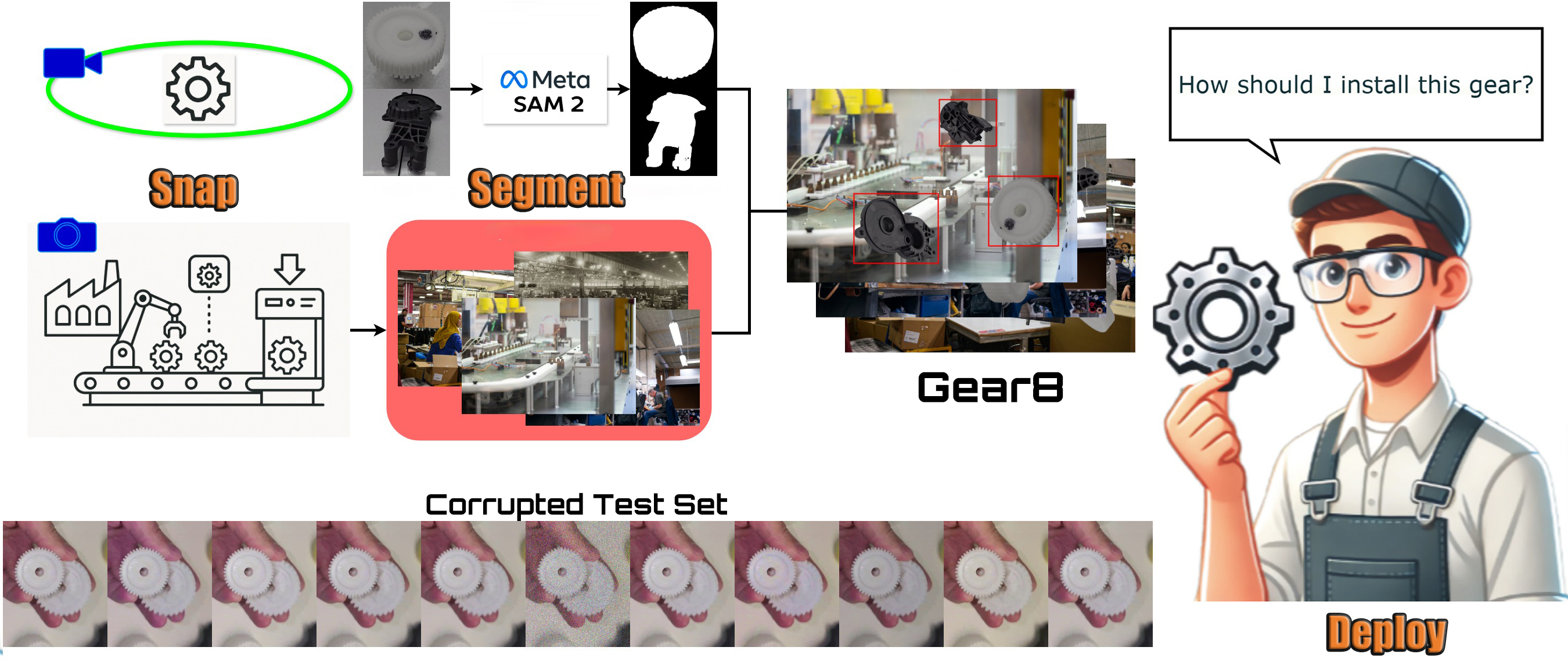}
}}
\caption{
Overview of our three-stage pipeline: \textbf{Snap}, \textbf{Segment}, and \textbf{Deploy}. 
In the \textit{Snap} stage, we collect multi-angle part videos and real-world background images from industrial environments. 
In the \textit{Segment} stage, we extract instance masks using SAM2~\cite{c58} and generate a synthetic dataset (Gear8) by compositing parts into factory scenes. In the \textit{Deploy} stage, a lightweight object detector trained on Gear8 is integrated into a wearable assistant system featuring speech input, visual recognition, knowledge retrieval, and audio feedback for interactive on-site guidance. The bottom row shows corrupted test images simulating real-world conditions for robustness evaluation.
}
\vskip-2ex
\label{fig:dataset_generation_process}
\end{figure}
However, the development of multimodal, on-device assistants for industrial use remains underexplored. Many existing solutions rely on cloud infrastructure, require extensive manual annotation, or fail to operate under practical constraints such as limited compute, offline usage, and strict data privacy~\cite{liu2024real,gharibvand2024cloud,mehta2024iar}.

We present a mobile-device-based industrial assistance system that leverages computer vision and Large Language Models (LLMs) to deliver real-time, context-aware guidance during assembly tasks. The system comprises a lightweight object detection module for part identification, a speech-to-text interface for natural language input, and a Retrieval-Augmented Generation (RAG) engine~\cite{c57} that generates part-specific responses, which are conveyed through text-to-speech output.
This end-to-end, on-device pipeline supports intuitive and semi-hands-free interaction, eliminating the need for cloud-based processing.
To facilitate training without manual annotations, we propose a fully automated data generation pipeline that synthesizes part-level training images using multi-view video captures and compositing over diverse industrial backgrounds.
Additionally, to improve the detector’s generalization under domain shift, we introduce a two-stage training approach termed Background-Agnostic Refinement (BAR), wherein the model is fine-tuned on plain-background object crops to prioritize object-centric features.

The system is modular, allowing components such as the vision model, knowledge base, or language model to be updated independently for new tasks or deployment settings. 

Bridging real-time object detection, vision-language interaction, and mobile industrial AI, our framework offers a deployable solution for contextual task assistance in constrained factory environments.

Extensive experiments on the Gear8 dataset, along with a structured user study, demonstrate the system’s effectiveness in improving detection robustness, task efficiency, and overall usability.

Our contributions are summarized as follows:
\begin{itemize}
    \item We design a real-time, multimodal assistant system for mobile industrial deployment, integrating visual detection, voice interaction, and semantic retrieval for hands-free part recognition and instruction.
    \item We propose a fully automated dataset generation pipeline that requires no manual annotation and supports diverse industrial backgrounds using consumer-grade equipment.
    \item We introduce a two-stage training strategy, Background-Agnostic Refinement (BAR), which improves detection robustness under domain shift without modifying model architecture.
\end{itemize}

\section{Related Work}

\subsection{Synthetic Data Generation for Vision Tasks}

Synthetic data has proven crucial for modern vision, initially in object detection and now across segmentation, pose estimation, and beyond. Early work used 3D renders: Peng et al.~\cite{c13} fine‑tuned networks pre‑trained on CAD images with minimal real data for PASCAL VOC. Game‑engine datasets like “Playing for Data”/“Playing for Benchmarks”~\cite{c14,c15} and Virtual KITTI~\cite{c16} provided pixel‑perfect urban scenes, while URSA~\cite{c17} and ProcSy~\cite{c18} scaled this to millions of driving images.

Domain randomization~\cite{c19,c20} addresses the sim‑to‑real gap by randomizing textures, lighting, and viewpoints. GAN‑based refinement improves realism: Nogues et al.~\cite{c21} and Hu et al.~\cite{c22} apply CycleGAN variants to industrial parts; Lin et al.~\cite{c23} use multi‑task GANs for traffic‑sign detection. BigDatasetGAN~\cite{c24} synthesizes pixel‑annotated ImageNet, and InstaGen~\cite{c25} employs diffusion models for open‑vocabulary detection. Copy‑paste augmentation remains simple yet effective~\cite{c26}. Synthetic data has also enabled progress in diverse tasks such as urban-scene parsing~\cite{c27}, remote-sensing detection~\cite{c28}, and medical or robotic vision. Moreover, domain-adaptive detectors~\cite{c29} and self-supervised learners benefit significantly from combining synthetic and real data.

We build on the copy‑paste paradigm by combining factory‑specific background images with instance masks extracted from full‑view object captures, producing diverse, domain‑adapted synthetic images that strike a balance between simplicity and high realism for industrial inspection.


\subsection{Real-Time Egocentric Vision in Wearable Systems}

Assistive technology~\cite{jiang2025atbench,liu2023open}, especially real-time egocentric vision~\cite{c48,c49}, has become increasingly relevant with the rise of smart glasses and edge AI devices. Several studies~\cite{c30, c31, c39} have demonstrated the deployment of YOLO-based detectors on wearable platforms for tasks like wire harness inspection and redundant part detection in assembly lines. Other works~\cite{c40, c41, c42, c43} extend this idea to maintenance and defect localization, leveraging egocentric views to support visual overlays and spatial guidance.

To address hardware limitations, researchers have proposed lightweight architectures~\cite{c33}, edge-cloud offloading~\cite{c32}, and wearable edge fusion~\cite{c35}, balancing latency and model performance. Some works further explore interaction modeling and long-term tracking~\cite{c44, c45}, supporting higher-level scene understanding in dynamic environments.

Assistive systems also benefit from real-time egocentric detection. Vision–tactile feedback~\cite{c46}, auditory guidance~\cite{c37}, and multi-modal hazard detection~\cite{c38} have shown promise in aiding the visually impaired or supporting safety in challenging environments. Additionally, MATERobot~\cite{c48} enables real-time material and object recognition with tactile feedback to improve environmental awareness for visually impaired individuals. Recent efforts such as ObjectFinder~\cite{c49} integrate open-vocabulary object detection and navigation to support interactive object search for blind users. Moreover, vision-language systems like EgoVLP~\cite{c50} and EdgeVL~\cite{c51} begin to bridge perception and semantic reasoning, though on-device deployment remains constrained. Notably, Vinci~\cite{c52} introduces a portable egocentric vision-language assistant capable of real-time scene understanding and user interaction on wearable devices.

Unlike prior work relying on proprietary hardware or cloud infrastructure, our system emphasizes low-cost, edge-deployable design for real-time multimodal perception and feedback in practical industrial workflows.

\section{Dataset Construction}
\label{sec:dataset}

Acquiring large-scale labeled datasets in industrial settings poses significant challenges due to factors such as strict confidentiality policies that limit third-party access and the lack of specialized annotation resources or infrastructure in small- and medium-sized enterprises.
In light of these constraints, it becomes essential to develop a data generation framework that empowers non-expert factory staff to efficiently produce high-quality visual datasets using readily available consumer-grade devices, without the need for advanced technical training.
To meet this objective, we introduce a synthetic dataset of industrial parts specifically designed to reflect the conditions and requirements of real-world deployment. This section outlines our methodology for data capture, the synthetic generation pipeline, and the key properties of the resulting dataset.
\subsection{Background Collection}
\label{dataset:background}

Due to restricted factory access, we selected 12 high-resolution, license-free industrial background images from public sources. Although this introduces a domain gap, it imposes a stricter evaluation setting and may underestimate real deployment performance.

\subsection{Multi-View Part Capture}
\label{dataset:object}

To simulate realistic conditions, we captured multi-view video sequences of each part using smartphones, without controlled environments or specialized equipment. The intentional use of different imaging devices from those in deployment introduces a domain gap, enabling robustness evaluation under suboptimal conditions.



From each video sequence $\{F_t\}_{t=1}^T$, a subset of frames is uniformly sampled across diverse viewpoints to maximize coverage while minimizing redundancy. These frames are then used for subsequent instance segmentation and synthetic data generation.

\subsection{Synthetic Composition} 
\label{dataset:composition}

To generate training samples, we first extract instance masks from each video frame $F_t$ using the SAM2~\cite{c58} model with center point guidance:
\begin{equation}
M_t = \mathrm{SAM_2}(F_t, c_t),
\label{eq:sam_mask}
\end{equation}
where $c_t$ denotes the center of the target object.

We then synthesize composite images by randomly pasting $k \in [3,5]$ object masks onto a selected background image $B$. Each mask undergoes random affine transformations:
\begin{equation}
m_i' = \mathrm{Affine}(m_i; \theta_i), \quad \theta_i = \{\text{scale}, \text{rotation}\}.
\label{eq:affine_transform}
\end{equation}
To ensure realism, we constrain the pairwise IoU between pasted instances to be below 0.5 and limit the number of instances per category to two per image. Object categories are sampled uniformly to ensure class balance.

The final composite image is defined as:
\begin{equation}
I = B \oplus \{m_1', \ldots, m_k'\}.
\label{eq:composite}
\end{equation}

This composition strategy promotes diversity in object appearance, position, and category, supporting generalizable and robust model training.

\subsection{Robustness Evaluation via Corruption Simulation}
\label{dataset:robustness}
To assess model robustness under deployment constraints, we construct a corrupted test set used exclusively for evaluation. We consider ten types of common perturbations, categorized into blur (motion blur, Gaussian blur), noise (Gaussian noise, ISO noise), and color or lighting variations (HSV modification, color shift, brightness adjustment, and contrast change). 


Each clean test image $I$ is corrupted by applying $c_i \in \mathcal{C}$ at a fixed, moderate severity $\alpha_i$:
\begin{equation}
I'_i = c_i(I; \alpha_i),
\label{eq:corruption}
\end{equation}
resulting in ten distinct corrupted versions per image, each reflecting a specific type of degradation.

Overall, our dataset construction strategy enables scalable, annotation-free training while ensuring diversity, realism, and robustness evaluation. The intentional domain gap between training and deployment conditions further validates the practical applicability of our approach to real-world industrial environments.

\section{Methodology}
\label{sec:method}

\begin{figure}[thpb]
      \centering
      \framebox{\parbox{3in}{
      \includegraphics[width=1.0\linewidth]{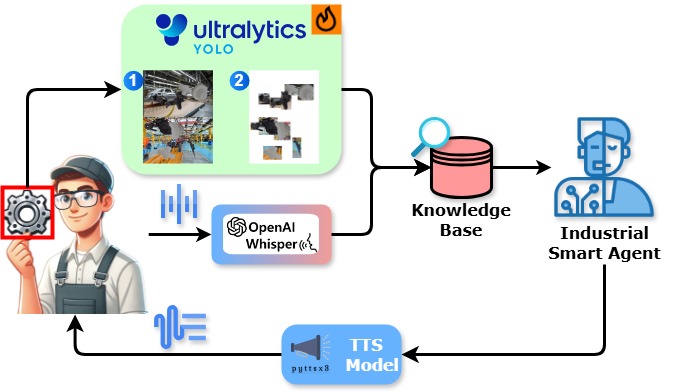}
}}

      \caption{System pipeline of the proposed industrial assistant integrates YOLO~\cite{khanam2024yolov11}-based detection, Whisper speech recognition, structured knowledge retrieval, and audio feedback. The detector is trained via a two-stage strategy with Background-Agnostic Refinement (BAR) to enhance robustness to domain shifts.}
      \vskip-2ex
      \label{Pipeline}
   \end{figure}
   
Our proposed industrial smart assistant system is designed to support real-time part recognition and multimodal user interaction under constrained computational environments. The pipeline consists of two main modules: a lightweight object detection framework enhanced by two-stage refinement, and a retrieval-augmented multimodal interaction system integrating visual and speech inputs.

\subsection{Background-Agnostic Refinement (BAR)}
\label{methodology:fadc}

To improve the generalization ability of lightweight object detectors under domain shifts, we adopt a two-stage training strategy called Background-Agnostic Refinement (BAR).

In the first stage, the detector is trained conventionally on the synthetic dataset containing diverse background conditions. After reaching high validation performance, we proceed to the second stage: the trained detector is re-applied to the original training set to extract high-confidence predictions. Each detected bounding box $b$ is cropped from its original image and placed onto a clean white canvas, eliminating background noise while preserving part-specific visual cues.

The resulting purified pseudo-labeled dataset emphasizes object-intrinsic features and reduces overfitting to contextual artifacts. We fine-tune the detector on this refined set, which empirically enhances robustness and improves detection stability across varying deployment environments.

\subsection{Retrieval-Augmented Multimodal Interaction}
\label{method:rag}

To enable intuitive user interaction with minimal manual effort, we design a retrieval-augmented multimodal system based on smart glasses. The pipeline consists of three modules:

\subsubsection{Knowledge Base Construction}
To enable the smart assistant to provide part-specific internal knowledge required in industrial scenarios, we construct a structured knowledge base covering all relevant components.

For each labeled part $x \in \mathcal{D}$, a semantic embedding $\mathbf{v}_x$ is computed using a pretrained SentenceTransformer~\cite{c60}:
\begin{equation}
\mathbf{v}_x = f_{\text{enc}}(x),
\label{eq:embedding}
\end{equation}
where $f_{\text{enc}}(\cdot)$ denotes the encoder function.

All embeddings are indexed using FAISS~\cite{c61}, an efficient library for similarity search. In our setting, FAISS constructs a flat L2 index to enable rapid nearest-neighbor retrieval based on Euclidean distance in the semantic space.

\subsubsection{Query Acquisition}
When the user triggers a query through a button press, the system continuously captures video frames $\{I_1, \ldots, I_n\}$ from the wearable device. To ensure that the user's field of view is properly aligned with the target object, detection is performed on each frame individually, and only when valid detections are observed over $N$ consecutive frames does the system proceed to the next step.

The final frame $I_n$ is used for subsequent processing. It is passed through our object detector, producing a set of object predictions. To improve detection robustness, we perform multi-frame fusion on the buffered detections across the last $N$ frames, aggregating predictions based on label consistency and bounding box IoU. For detections with the same label and IoU greater than a predefined threshold $\tau$, the bounding boxes are merged via confidence-weighted averaging:
\begin{equation}
b' = \frac{\sum_{i=1}^{k} w_i \, b_i}{\sum_{i=1}^{k} w_i},
\label{eq:box_fusion}
\end{equation}
where $b_i$ and $w_i$ denote the bounding box and its associated confidence score, respectively.
After fusion, redundant detections with high spatial overlap and identical labels are suppressed using a duplicate box filtering strategy, ensuring that each object is represented only once. The resulting fused and filtered detections are then passed through DepthAnything~\cite{c57} to infer pixel-wise depth maps.

The fused detections are then sorted in ascending order of depth, prioritizing closer objects for subsequent retrieval and interaction. Finally, the top-$K$ nearest objects (with $K=3$) are selected to construct the query candidates.

Formally, the output of this phase is:
\begin{equation}
\{(l_j, b_j, d_j)\}_{j=1}^{K},
\label{eq:yolo_depth_selection}
\end{equation}
where $l_j$ denotes the object label, $b_j$ the bounding box, and $d_j$ the estimated depth for the $j$-th selected object.

\subsubsection{Semantic Retrieval and Knowledge-Augmented Response Generation}
\label{method:retrieval_response}

During runtime, after the top-$K$ detected objects $\{(l_j, b_j, d_j)\}_{j=1}^{K}$ are selected based on depth ranking, each label $l_j$ is embedded into a vector representation $\mathbf{v}_{l_j}$ using a pretrained SentenceTransformer. These embeddings are used to query the FAISS index to retrieve the nearest database entries $x^*$, providing part-specific context $C$ for response generation.

Given the user's transcribed query $q$ and the retrieved context $C$, a language model (LLM) generates a natural language response tailored to the user's question: \begin{equation} \text{Answer} = \mathrm{LLM}(q, C). \label{eq:rag_answer} \end{equation} The generated answer is synthesized into speech via a text-to-speech (TTS) engine and played back to the user through the wearable device, enabling hands-free interaction.

\section{Experiments} 

\subsection{Dataset}
We construct the \textit{Gear8} dataset following the pipeline in Section~\ref{methodology:fadc}. It consists of two visually similar first-stage reduction gearboxes, each containing two gear types and a pair of cover and housing components.
To synthesize the training data, we composited segmented parts onto 12 industrial backgrounds sourced from license-free image platforms, selected to cover diverse industrial contexts. This process yields 4,000 training images and 1,000 validation images.
The test set comprises 196 real-world images capturing actual assembly scenarios and background environments. To simulate deployment conditions in practical systems, we further applied 10 types of realistic visual corruptions, resulting in an augmented test set of 2,156 images.

\subsection{Implementation Details}
All experiments are conducted using PyTorch 2.6.0 with CUDA 12.4. The YOLOv11n~\cite{khanam2024yolov11} model is fine-tuned on the \textit{Gear8} dataset, with second-stage training triggered once validation mAP50 exceeds 95.0. Multi-frame predictions are aggregated using confidence-weighted IoU merging. We set the detection confidence threshold to 0.4 and use a minimum of 5 consecutive frames with valid detections to confirm object presence. IoU threshold for merging is set to 0.5, with a minimum vote count of 3. Depth estimation is performed using the DepthAnything (vitl14)~\cite{c59} model. Semantic embeddings are generated using SentenceTransformer (all-MiniLM-L6-v2)~\cite{c60} and queried via FAISS~\cite{c61} over a structured component database. A local Phi-3-mini-4k-instruct~\cite{c62} language model is used for dialogue generation. Whisper-small~\cite{c56} handles speech recognition, and pyttsx3~\cite{c63} is used for text-to-speech synthesis. Voice input is recorded with a sample rate of 16kHz and a duration of 8 seconds. Real-time interaction is supported through depth-aware sorting of detected components and response synthesis based on retrieved metadata. 

\subsection{Results on Object Detection}

Table~\ref{tab:yolov11_results} presents the performance comparison under different combinations of training strategies and inference augmentations. The baseline model, YOLOv11n, shows limited accuracy due to its ultra-lightweight architecture. Applying Background-Agnostic Refinement (BAR) during training leads to consistent improvements across all metrics, particularly mAP@0.5:0.95, which rises from 0.06 to 0.17. This improvement stems from BAR's ability to decouple foreground object learning from background distractions. By refining the model with synthetic variations of the same objects on diverse backgrounds, BAR encourages the detector to focus on intrinsic object features rather than context-dependent cues, leading to better generalization under domain shifts.

Among inference-time methods, Test-Time Augmentation (TTA) yields the greatest gain. TTA aggregates predictions from spatially transformed inputs to improve robustness against scale and localization variations. When combined with BAR, it improves mAP@0.5 to 0.30 and mAP@0.5:0.95 to 0.20, demonstrating stronger generalization across object sizes and configurations.

\begin{table}[ht]
\centering
\caption{Performance comparison of different training strategies and inference augmentations}
\label{tab:yolov11_results}
\begin{tabular}{lcccc}
\hline
\textbf{Method} & \textbf{mP} & \textbf{mR} & \textbf{mAP@0.5} & \textbf{mAP@0.5:0.95} \\
\hline
YOLOv11n (Baseline) & 0.24 & 0.21 & 0.20 & 0.06 \\
w/ BAR & 0.28 & 0.25 & 0.26 & 0.17 \\
w/ BAR+SAHI & 0.23 & 0.23 & 0.20 & 0.06 \\
w/ BAR+TTA  & 0.35 & 0.26 & 0.30 & 0.20 \\
w/ BAR+TTA+SAHI & 0.31 & 0.26 & 0.25 & 0.16 \\
\hline
\end{tabular}
\end{table}

In contrast, SAHI~\cite{c54}, which slices input images into overlapping tiles for fine-grained detection, shows minimal or even negative effect. This is likely because our dataset primarily consists of small, well-centered objects in uncluttered settings. Tiling fragments the global context and introduces redundant detections, which ultra-lightweight models like YOLOv11n~\cite{khanam2024yolov11} are less capable of reconciling effectively. Notably, combining TTA and SAHI leads to a drop in performance compared to TTA alone, suggesting that the spatial fragmentation from SAHI interferes with the consistency benefits introduced by TTA.

Overall, BAR and TTA complement each other well in our setting, while SAHI proves less effective due to the dataset characteristics and model capacity. Based on this evaluation, we deploy the BAR+TTA-enhanced model in our industrial smart assistant system to ensure accurate, real-time performance under deployment constraints.

\subsection{Qualitative Results}
\begin{figure*}[thpb]
\centering
\includegraphics[width=0.98\textwidth]{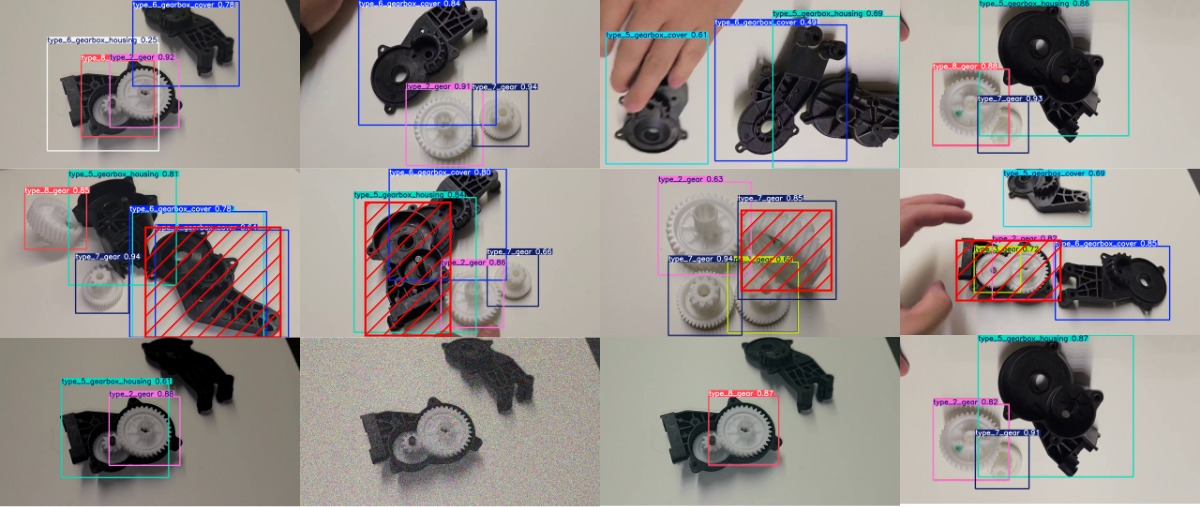}
\vskip-2ex
\caption{Qualitative detection results on the Gear8 dataset. The first row shows successful object detection cases on the uncorrupted test set, where all components are correctly identified. The second row displays failure cases on the same test set, where missed or incorrect detections are explicitly highlighted using red bounding boxes with hatched fill. The third row illustrates corruption-induced failures, as defined in Section~\ref{dataset:robustness}. All results are visualized using a confidence threshold of 0.25.}
\vskip-5ex
\label{fig:quantitative}
\end{figure*}

To further evaluate the performance of our framework, we present qualitative detection results for the object detection module under both clean and corrupted conditions, as shown in Figure~\ref{fig:quantitative}.

In the first row, accurate detections on uncorrupted test images demonstrate that the object detector is capable of handling real-world challenges such as partial occlusions and high inter-class similarity, indicating its ability to perform fine-grained recognition under clean conditions.

In contrast, the second row illustrates failure cases under clean conditions. Columns one to three show misclassifications of similar-looking parts—such as different versions of the gearbox cover, gearbox housing, and gears—particularly under specific viewpoints. These errors may be attributed to the limited capacity of the lightweight detector employed in our system, which may struggle to resolve subtle visual differences. Column four further demonstrates a failure under severe occlusion, where critical part features are no longer visible, leading to missed detections.

The third row depicts corruption-induced failures (as defined in Section~\ref{dataset:robustness}). Although the corresponding clean images were correctly detected, moderate perturbations lead to a notable drop in performance. In particular, ISO noise in column two results in complete detection failure, while imperceptible corruptions in column four cause misclassification into similar-looking parts. These observations confirm the sensitivity of the detection module to input distribution shifts and further validate the need for robustness-aware evaluation in real-world deployment settings.

\subsection{User Study}
Following preliminary experiments on object detection using our constructed dataset, we further evaluate the system in real-world scenarios with a focus on user experience in wearable settings. Given that usability and interaction quality are critical for practical deployment, we conduct a structured user study consisting of task-based evaluations and a post-task questionnaire to assess system effectiveness, usability, and user satisfaction.
\subsubsection*{Organization
}
To evaluate the effectiveness and usability of the system, we conducted a user study involving 9 participants (6 male, 3 female) in real-world assembly scenarios. For the evaluation, we selected components from 8 different object categories present in the training set, with varying instance counts for each class. Each participant interacted with the system for 30 minutes. To compare performance with and without system assistance, participants were asked to complete two rounds of part assembly: one without any system support, and another strictly following the system's step-by-step guidance. The time taken for each round was recorded separately. During the assisted phase, participants could also interact with the system to inquire about part-specific knowledge. After both rounds, all participants completed an anonymous questionnaire session. This included the NASA Task Load Index (NASA-TLX)~\cite{hart1988development} to assess cognitive workload, along with six custom-designed questions targeting the perceived usefulness, clarity, and reliability of our system. 

\subsubsection*{Performance}

To assess system performance from a user perspective, we designed six custom questions covering object recognition, instruction accuracy, task guidance, efficiency, voice clarity, and speech recognition. Participants rated each item on a 10-point Likert scale, with higher scores indicating better experience. As shown in Figure~\ref{fig:user_study}, the system received high ratings in task-related aspects, including \textit{Task Guidance} and \textit{Efficiency} (both 8.0), as well as in \textit{Voice Clarity} and \textit{Speech Recognition} (8.0). \textit{Instruction Accuracy} also performed well (7.0), suggesting consistent and understandable system output. The lower score in \textit{Object Recognition} (6.0) is attributed to the use of a lightweight YOLOv11n model (Table~\ref{tab:yolov11_results}) with only 2.6M parameters, necessary for real-time performance on mobile devices. Additional domain gaps caused by resolution differences between training and deployment environments further impacted detection stability. Despite mitigation through multi-frame fusion, recognition remains constrained by computational limitations. These results indicate that the proposed system offers reliable and efficient assistance in real-world assembly tasks, with potential for further improvement as hardware capabilities advance.

\subsubsection*{Time Efficiency and Error Prevention}
\begin{figure*}[htbp]
  \centering
  \includegraphics[width=\textwidth,keepaspectratio]{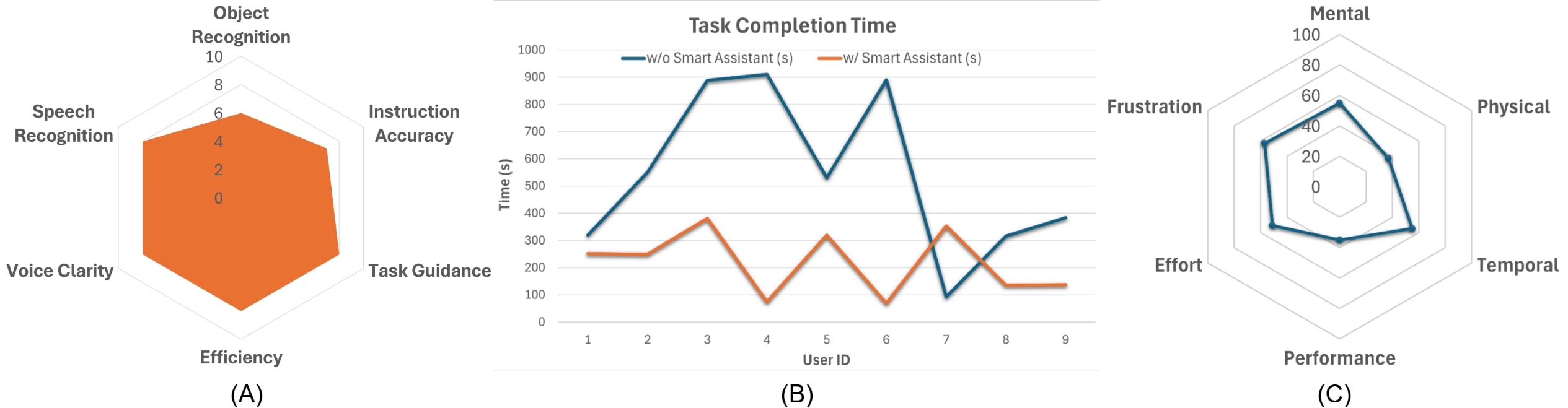}
  \vskip-2ex
  \caption{User-study results for the Smart Assistant system:  
           (a) average scores on the customized functionality-assessment questionnaire (0–10, higher is better);  
           (b) task-completion time with and without the Smart Assistant for nine participants;  
           (c) average NASA-TLX workload scores (0–100, lower is better).}
        \vskip-4ex
  \label{fig:user_study}
\end{figure*}

To evaluate the impact of the smart assistant on operational performance, we recorded task completion times for each participant with and without system support. As shown in Table~\ref{fig:user_study}, the average completion time decreased from 542.3\,s to 219.2\,s, demonstrating a substantial improvement in task efficiency. An exception was observed for participant ID~7, who completed the task in 93\,s unaided but took 354\,s with guidance. This likely reflects prior task familiarity, where the system's structured workflow introduced additional overhead. Such outliers are rare and do not affect the overall trend.

In addition to improved efficiency, the system also enhanced operational safety. In the unassisted condition, a total of 5 parts were damaged due to incorrect interference fits, resulting in surface abrasion on shaft components. No such incidents occurred under system guidance, indicating that the assistant effectively prevents critical assembly errors and reduces potential production costs.

\subsubsection*{Cognitive Load}
To evaluate the cognitive load introduced by our wearable system, we employed the NASA Task Load Index (NASA-TLX)~\cite{hart1988development}, a widely used measure of subjective workload. Participants rated six dimensions: Mental, Physical, Temporal Demand, Performance, Effort, and Frustration. Figure~\ref{fig:user_study} presents the average scores across all participants. The overall score was 48.0, close to the global median (range: 6.21–88.50, $N=1173$~\cite{c53}), indicating moderate workload. The system is effective without excessive strain. Among all factors, \textit{Frustration} (57) and \textit{Mental Demand} (55) were relatively high. Likely due to the effort required to interact with a multimodal system, especially when precise part names are needed. Occasional language model hallucinations, verbose outputs, or extended response times may also have contributed. In contrast, the lowest score was for \textit{Physical Demand} (37), confirming that the semi-hands-free interface minimizes physical effort. Overall, the system provides a positive user experience with acceptable cognitive load with potential for further improvement through interaction refinement.

\subsubsection*{Overall Rating}
Participants rated their likelihood to recommend the Smart Assistant system to others on a 0–10 scale. The average score was 6.11 (SD = 1.62), indicating moderate satisfaction. Individual scores ranged from 4 (lowest) to 8 (highest), highlighting variability in user perceptions and experiences.

\subsubsection*{Limitations}

The current system is subject to two main limitations. First, due to safety and privacy constraints in industrial environments, the system runs fully offline on resource-constrained wearable devices. This requires lightweight models to meet real-time constraints, limiting overall capacity and accuracy. Second, the devices used for data collection and deployment differ in resolution and sensor characteristics, introducing a domain gap that may affect detection performance during testing. These issues, however, are not fundamental limitations of the proposed pipeline. In practice, they can be addressed by aligning the deployment hardware with the data acquisition setup, and by integrating higher-performance servers within the factory's local network. Such improvements would allow the adoption of more capable models while preserving the modular structure and deployment flow of the system, leading to further gains in recognition accuracy and interaction quality.

\section{Conclusion}
We introduced a semi-hands-free smart assistant for industrial assembly, tailored for offline use, limited compute, and strict privacy. The system integrates lightweight object detection, depth estimation, and retrieval-augmented dialogue in a mobile-friendly pipeline. To avoid manual labeling, we developed an automated data generation process and a two-stage training method (BAR) to boost detection robustness.
Experiments on the Gear8 dataset show improved accuracy with high efficiency, while a user study confirmed gains in task speed, reduced part damage, and lower cognitive load. Despite current hardware and domain shift limitations, future updates with local servers and aligned data capture can address these. Our approach provides a scalable, privacy-preserving solution for intelligent industrial assistance.

\section*{ACKNOWLEDGMENT}
The project served to prepare the SFB 1574 Circular Factory for the Perpetual Product (project ID: 471687386), approved by the German Research Foundation (DFG, German Research Foundation). This work was supported in part by the SmartAge project sponsored by the Carl Zeiss Stiftung (P2019-01-003; 2021-2026), the MWK through the Cooperative Graduate School Accessibility through AI-based Assistive Technology (KATE) under Grant BW6-03, and in part by the BMBF through a fellowship within the IFI program of the German Academic Exchange Service (DAAD), in part by the HoreKA@KIT supercomputer partition, and in part by the National Natural Science Foundation of China (No. 62473139).

\bibliographystyle{unsrt}
\bibliography{bib.bib}

\end{document}